\documentclass[aps,prl,twocolumn,groupedaddress]{revtex4}

\usepackage{graphicx}
\usepackage{dcolumn}
\usepackage{bm}
\usepackage{color, soul}
\soulregister\cite7 %
\begin{document}

\title{Superconductivity in hexagonal Nb-Mo-Ru-Rh-Pd high-entropy alloys}

\date{\today}
\author{Bin Liu$^{1,2,3}$, Jifeng Wu$^{1,2,3}$, Yanwei Cui$^{2,3,4}$, Qinqing Zhu$^{1,2,3}$, Guorui Xiao$^{2,3,4}$, Siqi Wu$^{4}$, Guanghan Cao$^{4}$}
\author{Zhi Ren$^{2,3}$\footnote{Corresponding author: renzhi@westlake.edu.cn}}

\affiliation{$^{1}$Department of Physics, Fudan University, Shanghai 200433, P. R. China}
\affiliation{$^{2}$School of Science, Westlake University, 18 Shilongshan Road, Hangzhou 310064, P. R. China}
\affiliation{$^{3}$Institute of Natural Sciences, Westlake Institute for Advanced Study, 18 Shilongshan Road, Hangzhou 310064, P. R. China}
\affiliation{$^{4}$Department of Physics, Zhejiang University, Hangzhou 310027, P. R. China}

\begin{abstract}
We report the superconducting properties of new hexagonal Nb$_{10+2x}$Mo$_{35-x}$Ru$_{35-x}$Rh$_{10}$Pd$_{10}$ high-entropy alloys (HEAs) (0 $\leq$ $x$ $\leq$ 5). With increasing $x$, the superconducting transition temperature $T_{\rm c}$ shows a maximum of 6.19 K at $x$ = 2.5, while the zero-temperature upper critical field $B_{\rm c2}$(0) increases monotonically, reaching 8.3 T at $x$ = 5. For all $x$ values, the specific heat jump deviates from the Bardeen-Cooper-Schreiffer behavior. In addition, we show that $T_{\rm c}$ of these HEAs is not determined mainly by the density of states at the Fermi level and would be enhanced by lowering the valence electron concentration.
\end{abstract}

\maketitle
\maketitle

Recently, high-entropy alloys (HEAs) have received a lot of attention because of their superior properties compared with traditional alloys \cite{HEAreview1,HEAreview2,HEAreview3,HEAreview4,HEAreview5}.
These multicomponent systems contain five or more metallic elements, whose molar fractions vary between 5\% and 35\%.
This diversity of constituent elements results in high mixing entropy and strong atomic scale disorder.
Hence HEAs are often referred as metallic glasses on an ordered lattice and found to crystallize mostly in simple structures, including body-centered cubic (bcc) \cite{bccHEA1,bccHEA2,bccHEA3,bccHEA4,bccHEA5,bccHEA6}, face-centered cubic (fcc) \cite{fccHEA1,fccHEA2,fccHEA3,fccHEA4,fccHEA5} and hexagonal closed packing (hcp) \cite{hcpHEA1,hcpHEA2,hcpHEA3,hcpHEA4,hcpHEA5}.
Due to the chemical complexity, the properties of HEAs can be highly tunable, which provides a versatile platform to study the structure-property relationship of alloys in between the crystalline and glassy states \cite{cgstates}.

Superconductivity in HEAs was first discovered in the Ta-Nb-Hf-Zr-Ti system in 2014 \cite{HEASC}. Since then, a number of HEA superconductors have been discovered \cite{HEASCreivew} and most of them have the cubic structures with different space groups, such as bcc-type \cite{HEASC}, $\alpha$-Mn type \cite{alphaMn}, CsCl-type \cite{CsClHEASC}, and A15-type \cite{A15HEA}. In particular, a highest $T_{\rm c}$ of 10.2 K is achieved in the A15-type V$_{1.4}$Nb$_{1.4}$Mo$_{0.2}$Al$_{0.5}$Ga$_{0.5}$ HEA \cite{A15HEA}. By contrast, the hexagonal HEA superconductors have been rare, which is mainly due to the difficulty in forming hexagonal HEAs since most elements in the periodic table prefer a bcc or fcc structure \cite{hcpdiff}. The first one of this kind is the Re$_{0.56}$Nb$_{0.11}$Ti$_{0.11}$Zr$_{0.11}$Hf$_{0.11}$ HEA, which has a $T_{\rm c}$ of 4.4 K and a $B_{\rm c2}$(0) of 3.6 T \cite{hcpHEASC1}. Nevertheless, a hexagonal to bcc structural transition occurs by lowering the Re content. Later on, Lee \emph{et al.} reported the observation superconductivity in hexagonal Mo$_{23.75}$Re$_{23.75}$Ru$_{23.75}$Rh$_{23.75}$Ti$_{5}$ and Mo$_{22.5}$Re$_{22.5}$Ru$_{22.5}$Rh$_{22.5}$Ti$_{10}$ HEAs with $T_{\rm c}$ values of 3.6 K and 4.7 K, respectively, though no other superconducting properties have been characterized \cite{hcpHEASC2}.
In this context, the search for HEA superconductors with a hexagonal structure is highly desirable.

\begin{figure*}
\includegraphics*[width=15cm]{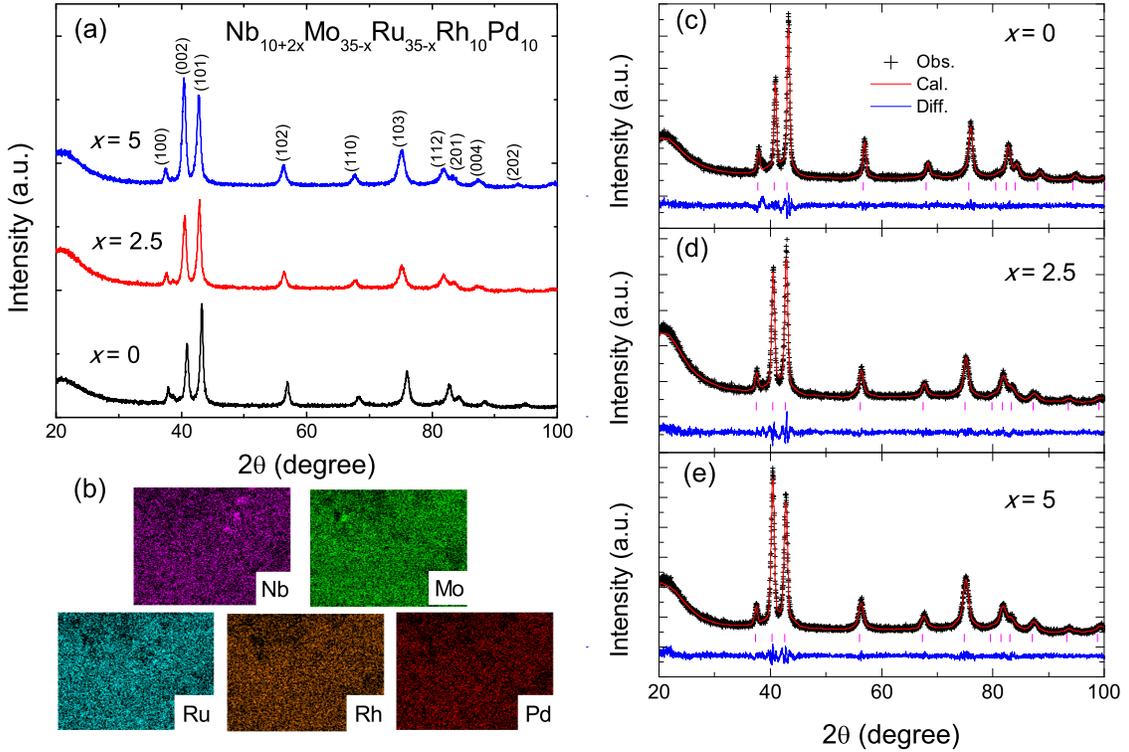}
\caption{
(a) Powder x-ray diffraction patterns for the series of Nb$_{10+2x}$Mo$_{35-x}$Ru$_{35-x}$Rh$_{10}$Pd$_{10}$ HEAs. The diffraction peaks for $x$ = 5 are indexed based on a hexagonal unit cell. (b) EDX elemental mapping of the Nb, Mo, Ru, Rh and Pd elements for the HEA with $x$ = 2.5. (c)-(e) Structural refinement profiles for the HEAs with $x$ = 0, 2.5, and 5, respectively. The refinement results are $R_{\rm wp}$ = 5.9\%, $R_{\rm p}$ = 4.6\%, goodness-of-fit (GOF) = 1.28 for $x$ = 0; $R_{\rm wp}$ = 5.6\%, $R_{\rm p}$ = 4.5\%, GOF = 1.23 for $x$ = 2.5; $R_{\rm wp}$ = 5.5\%, $R_{\rm p}$ = 4.4\%, GOF = 1.23 for $x$ = 5.
}
\label{fig1}
\end{figure*}
\begin{table*}
\caption{Atomic coordinates for the Nb$_{10+2x}$Mo$_{35-x}$Ru$_{35-x}$Rh$_{10}$Pd$_{10}$ HEAs.}
\renewcommand\arraystretch{1.3}
\begin{tabular}{p{5cm}<{\centering}p{0.8cm}<{\centering}p{0.8cm}<{\centering}p{0.8cm}<{\centering}p{0.8cm}<{\centering}p{8cm}<{\centering}}
\\
\hline 
   Atoms & site  &  $x$  & $y$ & $z$ & Occupancy \\

\hline 
Nb(1)/Mo(1)/Ru(1)/Rh(1)/Pd(1)							& 	 2$a$ 	 & 0	&0 & 0 & (0.10 + 0.02$x$)/(0.35-0.01$x$)/(0.35-0.01$x$)/0.10/0.10 	\\
Nb(2)/Mo(2)/Ru(2)/Rh(2)/Pd(2)							& 	  4$f$ 	 & 2/3 & 1/3 & 1/2 & (0.10 + 0.02$x$)/(0.35-0.01$x$)/(0.35-0.01$x$)/0.10/0.10		\\
\hline
\hline 
\end{tabular}
\label{Table3}
\end{table*}
In this paper, we present a study of the crystal structure and physical properties of the Nb$_{10+2x}$Mo$_{35-x}$Ru$_{35-x}$Rh$_{10}$Pd$_{10}$ HEAs for 0 $\leq$ $x$ $\leq$ 5, which are based on the quaternary Mo-Ru-Rh-Pd alloy \cite{MoReRuPd}. All these HEAs are shown to adopt the hcp structure and discovered to be bulk superconductors. A maximum $T_{\rm c}$ of 6.19 K and $B_{\rm c2}$(0) of 8.3 T are observed at $x$ = 2.5 and 5, respectively. The specific heat results suggest that these HEAs have a non BCS-like gap. Moreover, the $x$ dependence of the electronic specific coefficient differs from that of $T_{\rm c}$, implying that the density of states at the Fermi level is not the dominant factor in determining $T_{\rm c}$. A comparison of the valence electron concentration (VEC) dependence of $T_{\rm c}$ is made between the Nb-Mo-Ru-Rh-Pd HEAs and other hexagonal HEA superconductors, and its implication is discussed.

The Nb-Mo-Ru-Rh-Pd HEAs were prepared by the standard arc-melting method. High purity Nb (99.99\%), Mo (99.9\%), Ru (99.99\%), Rh (99.99\%), Pd (99.99\%) powders were mixed according to the stoichiometric ratios Nb:Mo:Ru:Rh:Pd = (10+2$x$):(35-$x$):(35-$x$):10:10 ($x$ = 0, 2.5 and 5) and pressured into pellets in an argon-filled glove box. Then the pellets were melted several times in an arc furnace under high-purity argon atmosphere, followed by rapid cooling on a water-chilled copper plate. The weight loss during the arc-melting process was found to be negligible. Since the resulting HEAs are ductile, they could not be ground into powders. Hence the x-ray diffraction (XRD) measurements were done on flat surface of the samples using a Bruker D8 Advance x-ray diffractometer with a monochromatic Cu-K$\alpha$ radiation at room temperature. The structural refinements were performed using the JANA2006 program \cite{JANA2006}. The chemical compositions of the HEAs were investigated by Hitachi field emission scanning electron microscope (SEM) equipped with an energy dispersive x-ray (EDX) spectrometer. The electrical resistivity was measured by a standard four-probe method. The resistivity and specific heat measurements were done in a Quantum Design Physical Property Measurement System (PPMS-9 Dynacool). The dc magnetization measurements were performed using a commercial SQUID magnetometer (MPMS3).

\begin{figure*}[t]
\includegraphics*[width=15cm]{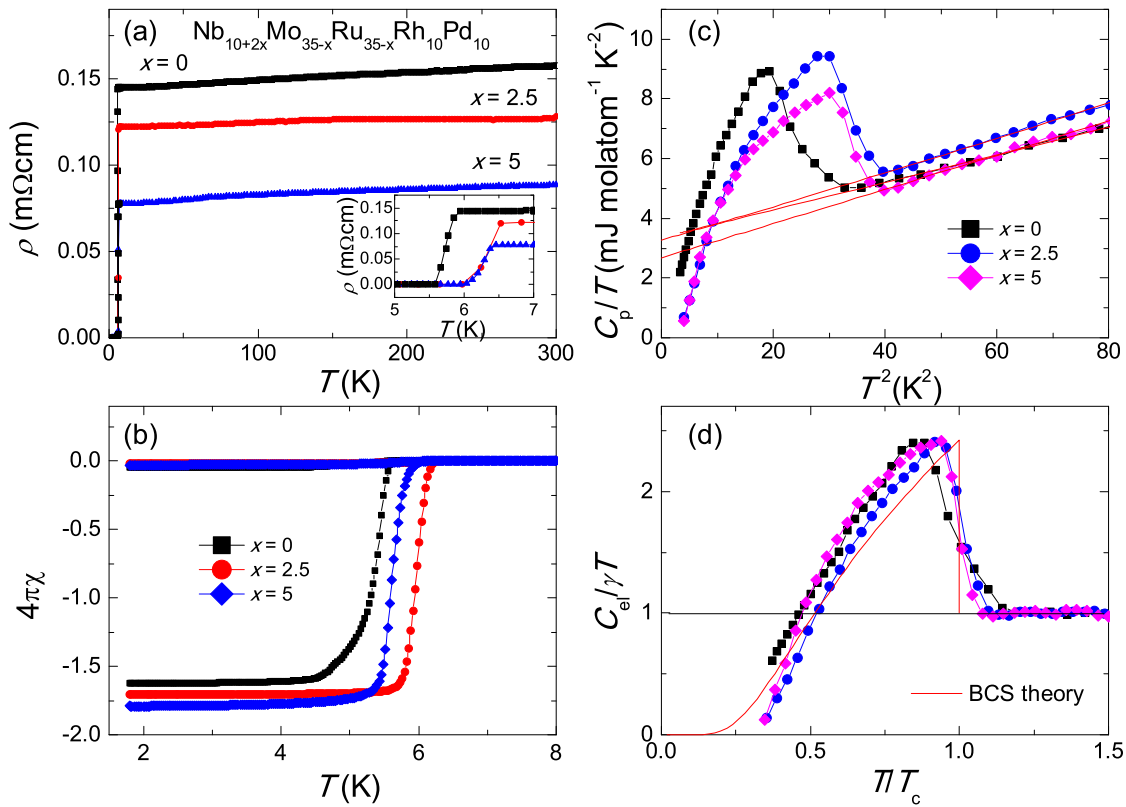}
\caption{
(a) Temperature dependence of resistivity for the series of Nb$_{10+2x}$Mo$_{35-x}$Ru$_{35-x}$Rh$_{10}$Pd$_{10}$ HEAs. The inset shows a zoom of the data near the resistive transitions.
(b) Temperature dependence of zero-field cooling and field cooling magnetic susceptibilities for the HEAs.
(c) Low temperature specific heat data for the HEAs plotted as $C_{\rm p}$/$T$ versus $T^{2}$.
The solid lines are Debye fits to the normal-state data.
(d) Normalized electronic specific heat plotted against $T$/$T_{\rm c}$ for these HEAs.
The solid line is the calculated curve from the BCS theory.
}
\label{fig1}
\end{figure*}
Figure 1(a) shows the XRD patterns for the series of Nb$_{10+2x}$Mo$_{35-x}$Ru$_{35-x}$Rh$_{10}$Pd$_{10}$ HEAs.
One can see that the patterns are very similar for these HEAs, and all the diffraction peaks can be well indexed on a hexagonal unit cell with the $P$$6_{3}$/$mmc$ space group.
The refined lattice parameters are $a$ = 2.749(1) {\AA}, $c$ = 4.423(1) {\AA}, $a$ = 2.773(1) {\AA}, $c$ = 4.474(1) {\AA} and $a$ = 2.773(1) {\AA}, $c$ = 4.468(1) for $x$ = 0, 2.5 and 5, respectively.
Clearly, both $a$- and $c$-axis tend to increase with the increase of Nb content, which is as expected since the atomic radius of Nb is larger than those of Mo and Ru \cite{radius}.
The chemical compositions as determined by EDX measurements are Nb$_{11.8}$Mo$_{31.3}$Ru$_{39.8}$Rh$_{7.1}$Pd$_{10}$, Nb$_{13.9}$Mo$_{32.3}$Ru$_{34.7}$Rh$_{9.1}$Pd$_{10}$, Nb$_{20.3}$Mo$_{28.2}$Ru$_{33.6}$Rh$_{7.6}$
Pd$_{10.3}$  for $x$ = 0, 2.5 and 5, respectively, which agree with the nominal ones within the experimental error.
Moreover, EDX elemental mapping reveals a uniform distribution of the constituent elements Nb, Mo, Ru, Rh, Pd, and an example for $x$ = 2.5 is shown in Fig. 1(b).
To determine the atomic arrangement in the hexagonal unit cell, structural refinements were performed on the XRD of these samples and the results are plotted in Fig. 1(c)-(e). In all cases,
it is found that the five elemental atoms are disorderly distributed on the two crystallographic sites (0, 0, 0) and (2/3, 1/3, 1/2), as can be seen in Table I. Furthermore, good agreements between the observed and calculated patterns are evidenced by the small $R_{\rm wp}$/$R_{\rm p}$ values and goodness-of-fits close to 1.2 (see Table II).
Taken together, these results indicate that these HEA samples are homogeneous random solid solutions.

The temperature dependence of resistivity ($\rho$) for the Nb$_{10+2x}$Mo$_{35-x}$Ru$_{35-x}$Rh$_{10}$Pd$_{10}$ HEAs is shown in Fig. 2(a).
These HEAs exhibits a weak metallic behavior below 300 K, and the residual resistivity ratio $\rho_{\rm 300 K}$/$\rho_{\rm 7 K}$ is only around 1.1.
This is reminiscent of that observed in many other HEAs \cite{alphaMn,CsClHEASC,A15HEA} and ascribed to the presence of strong atomic-scale disorder.
With increasing Nb content $x$, the $\rho$ magnitude decreases strongly, which is probably due to the increase in either carrier concentration or mobility.
As seen more clearly in the inset of Fig. 2(a), a rapid $\rho$ drop is observed at low temperature for all samples, indicative of a superconducting transition.
Concomitant with the resistive transition, the zero-field cooling magnetic susceptibility data display a strong diamagnetic response, as displayed in Fig. 2(b).
Furthermore, the onset of the diamagnetic transition coincides well with that of the zero resistivity.
This allows us to determine $T_{\rm c}$ to be 5.58 K, 6.19 K, and 6.10 K for $x$ = 0, 2.5, and 5, respectively.
At 1.8 K, the zero-field susceptibility data of these HEAs correspond to shielding fractions more than 150\%.
Although the demagnetization effect is not taken into consideration,
these large values strongly suggests that the observed superconductivity is a bulk effect.

\begin{figure}[t]
\includegraphics*[width=8.3cm]{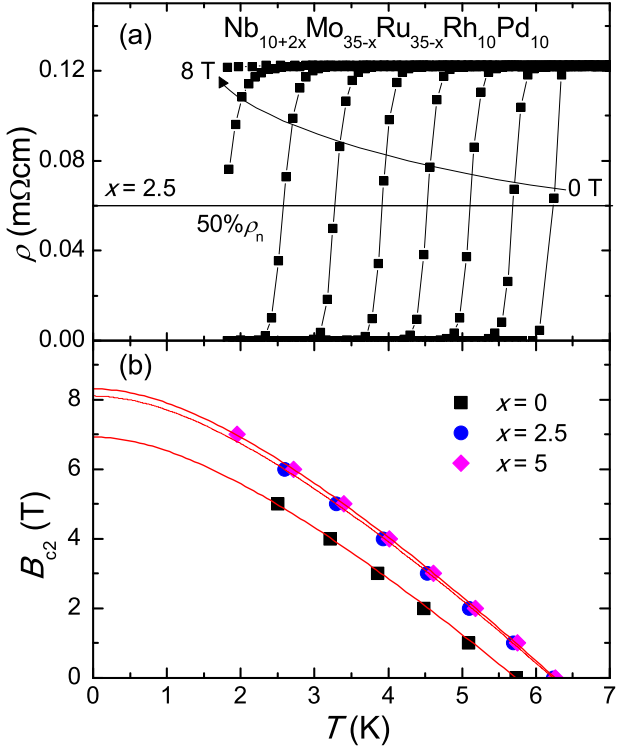}
\caption{
(a) Temperature dependence of resistivity under various magnetic fields up to 8 T for the Nb$_{10+2x}$Mo$_{35-x}$Ru$_{35-x}$Rh$_{10}$Pd$_{10}$ HEA with $x$ = 2.5. The arrow marks the increasing field direction, and the field increment for each curve is 1 T. The horizontal line represents the level corresponding to 50\% of the normal-state resistivity just above $T_{\rm c}$.
(b) Temperature dependence of upper critical fields for the Nb$_{10+2x}$Mo$_{35-x}$Ru$_{35-x}$Rh$_{10}$Pd$_{10}$ HEAs.
The solid lines are fits from the WHH model to the data.
}
\label{fig1}
\end{figure}

The bulk nature of superconductivity is further confirmed by the specific-heat ($C_{\rm p}$) measurements.
As can be seen from Fig. 2(c), a clear $C_{\rm p}$ jump is detected for each of the HEAs around $T_{\rm c}$.
Following the standard practice \cite{A15HEA}, the normal-state $C_{\rm p}$ data are analyzed by the Debye model $C_{\rm p}$/$T$ = $\gamma$ + $\beta_{3}$$T^{2}$ + $\beta_{5}$$T^{4}$, where $\gamma$ and $\beta_{i}$ ($i$ = 3, 5) are the electronic and phonon specific heat coefficients, respectively. Once $\beta_{3}$ is known, the Debye temperature $\Theta_{\rm D}$ is calculated as $\Theta_{\rm D}$ = (12$\pi$$^{4}$$R$$/5\beta$)$^{1/3}$, where $R$ is the molar gas constant 8.314 J mol$^{-1}$ K$^{-2}$. These analyses give $\gamma$ = 3.35 mJ molatom$^{-1}$ K$^{-2}$, $\Theta_{\rm D}$ = 348 K, $\gamma$ = 3.27 mJ molatom$^{-1}$ K$^{-2}$, $\Theta_{\rm D}$ = 328 K, and $\gamma$ = 2.68 mJ molatom$^{-1}$ K$^{-2}$, $\Theta_{\rm D}$ = 324 K for $x$ = 0, 2.5 and 5, respectively. After subtraction of the phononic contribution, the normalized electronic specific heat $C_{\rm el}$/$\gamma$$T$ is plotted against the reduced temperature $T$/$T_{\rm c}$ in Fig. 2(d). One can see that the $C_{\rm el}$/$\gamma$$T$ jumps for all these HEAs are very close to the BCS value of 1.43 \cite{BCS}, but their temperature dependencies deviate clearly from the behavior expected from the BCS theory \cite{BCS}. Note that, for all these HEAs, the extrapolation of $C_{\rm el}$/$\gamma$$T$ data to 0 K yields a negative value, pointing to a full gapped superconducting state.
Hence the deviation from the BCS behavior could be due to the presence of multiple gaps.
Nevertheless, the fitting with the multigap model does not give satisfactory result. This is probably due to that the measurement temperature is not low enough, or that the effect of strong disorder is not included in the model.
Regarding the latter, it has been shown theoretically that the size of $C_{\rm p}$ jump for two-gap superconductors depends critically on the level of disorder \cite{disordertwogap}. In this respect, further experimental and theoretical studies in future are necessary to achieve a proper understanding of the $C_{\rm p}$ data of these HEAs.
Assuming a phonon mediated pairing mechanism, the electron-phonon coupling constant $\lambda_{\rm ep}$ for these HEAs can be estimated using the inverted McMillan formula \cite{Mcmillan},
\begin{equation}
\lambda_{\rm ep} = \frac{1.04 + \mu^{\ast} \rm ln(\Theta_{\rm D}/1.45\emph{T}_{\rm c})}{(1 - 0.62\mu^{\ast})\rm ln(\Theta_{\rm D}/1.45\emph{T}_{\rm c}) - 1.04},
\end{equation}
where $\mu^{\ast}$ = 0.13 is the Coulomb repulsion pseudopotential. Thus $\lambda_{\rm ep}$ values of 0.63, 0.66, and 0.66 are found for $x$ = 0, 2.5 and 5, respectively, suggesting that the Nb$_{10+2x}$Mo$_{35-x}$Ru$_{35-x}$Rh$_{10}$Pd$_{10}$ HEAs are intermediate coupling superconductors.

\begin{figure}[t]
\includegraphics*[width=8.7cm]{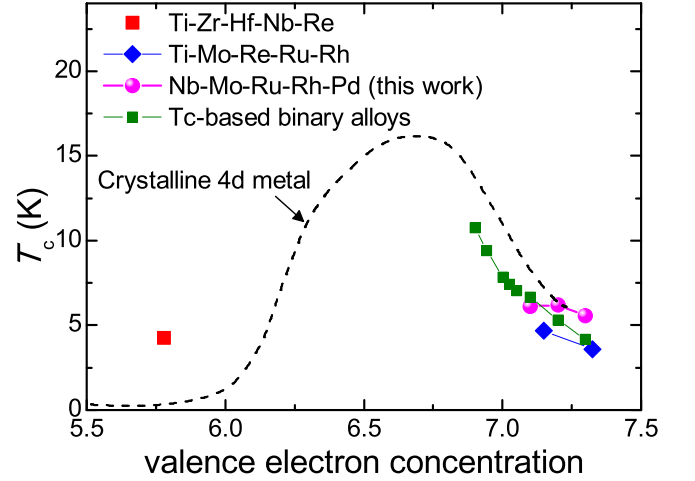}
\caption{
Comparison of the VEC dependence of $T_{\rm c}$ between the hexagonal HEA superconductors. The data for crystalline 4$d$ metals \cite{MTrule} and Tc-based binary hcp alloys \cite{Tcbasedalloys} are also included for reference.
}
\label{fig1}
\end{figure}
The upper critical fields ($B_{\rm c2}$) of the Nb$_{10+2x}$Mo$_{35-x}$Ru$_{35-x}$Rh$_{10}$Pd$_{10}$ HEAs were determined by resistivity measurements.
As an example, Fig. 3(a) shows the $\rho$($T$) curves under various magnetic fields up to 8 T for $x$ = 2.5.
With increasing field, the resistive transition is gradually suppressed to lower temperature.
For each curve, the resistive $T_{c}$ is defined as the temperature corresponding to midpoint of the $\rho$ drop.
The resulting temperature dependencies of $B_{\rm c2}$ for these HEAs are shown in Fig. 3(b).
To obtain the zero temperature upper critical field $B_{\rm c2}$(0), the data are extrapolated using the Wathamer-Helfand-Hohenberg theory \cite{WHH}.
This gives $B_{\rm c2}$(0) = 6.9 T, 8.1 T, and 8.3 T for $x$ = 0, 2.5 and 5, respectively.
Since the Ginzburg-Landua (GL) coherence length $\xi_{\rm GL}$ can be calculated by the equation $\xi_{\rm GL}$ = $\sqrt{\Phi_{0}/2\pi B_{\rm c2}(0)}$, where $\Phi_{0}$ = 2.07 $\times$ 10$^{-15}$ Wb is the flux quantum,
we obtain $\xi_{\rm GL}$ = 6.9 nm, 6.4 nm, and 6.3 nm for these HEAs.

\begin{table*}
\caption{Chemical compositions, lattice and physical parameters of the Nb$_{10+2x}$Mo$_{35-x}$Ru$_{35-x}$Rh$_{10}$Pd$_{10}$ HEAs.}
\renewcommand\arraystretch{1.3}
\begin{tabular}{p{3.5cm}<{\centering}p{4.5cm}<{\centering}p{4.5cm}<{\centering}p{4.5cm}<{\centering}p{0.8cm}}
\\
\hline 
   Parameter  &  $x$ = 0  & $x$ = 2.5  & $x$ = 5 \\
\hline 
Measured composition & Nb$_{11.8}$Mo$_{31.3}$Ru$_{39.8}$Rh$_{7.1}$Pd$_{10}$  & 	Nb$_{13.9}$Mo$_{32.3}$Ru$_{34.7}$Rh$_{9.1}$Pd$_{10}$   	 & Nb$_{20.3}$Mo$_{28.2}$Ru$_{33.6}$Rh$_{7.6}$Pd$_{10.3}$	 \\
$a$ ({\AA}) & 2.749(1) 	&  2.773(1)      & 2.773(1)   \\
$c$ ({\AA})& 4.423(1) & 	4.474(1)  	 & 4.468(1) 	\\
$R_{\rm wp}$ & 5.9\% &	5.6\%  	 & 5.5\% 	\\
$R_{\rm p}$ & 4.6\% &	4.5\% 	 & 4.4\% 	\\
GOF & 1.28&	1.23 	 & 1.23	\\
$T_{\rm c}$ (K) &5.58  & 6.19   	 & 6.10  	\\
$\gamma$ (mJ molatom$^{-1}$ K$^{-2}$) & 3.35& 	3.27  	 & 2.68 	\\
$\Theta_{\rm D}$ (K) & 348 & 	328  	 & 324 	\\
$\lambda_{\rm ep}$ & 0.63& 	0.66 	 & 0.66 	\\
$B_{\rm c2}$(0) (T)& 6.9& 8.1  	 & 8.3	\\
$\xi_{\rm GL}$ (nm)& 6.9 & 	6.4  	 & 6.3 	\\
\hline
\hline 
\end{tabular}
\label{Table3}
\end{table*}
From the above results, which are summarized in Table I,
one can see that the Nb$_{10+2x}$Mo$_{35-x}$Ru$_{35-x}$Rh$_{10}$Pd$_{10}$ HEAs show a maximum $T_{\rm c}$ of 6.19 K at $x$ = 2.5 and a maximum $B_{\rm c2}$(0) of 8.3 T at $x$ = 5. While both these values are the highest among hexagonal HEA superconductors, it is noted that the maximum $T_{\rm c}$ is not correlated with the maximum in $\gamma$.
Instead, the maximum $\gamma$ is observed for the HEA with $x$ = 0, which has the lowest $T_{\rm c}$.
Nevertheless, there appears a correspondence between $\lambda_{\rm ep}$ and $T_{\rm c}$, namely, the larger $\lambda_{\rm ep}$ the higher $T_{\rm c}$.
For the hexagonal Ti-Zr-Hf-Nb-Re HEA superconductor with a lower $\lambda_{\rm ep}$ = 0.57 \cite{hcpHEASC1}, its $T_{\rm c}$ of 4.4 K is indeed lower than the Nb$_{10+2x}$Mo$_{35-x}$Ru$_{35-x}$Rh$_{10}$Pd$_{10}$ HEAs.
These results suggest that electron phonon coupling strength plays a more important role than the density of states at the Fermi level in determining $T_{\rm c}$ in hexgonal HEAs.

Finally, in Fig. 4, we compare the VEC dependence of $T_{c}$ for the Nb$_{10+2x}$Mo$_{35-x}$Ru$_{35-x}$Rh$_{10}$Pd$_{10}$ HEAs with other hexagonal HEA superconductors. As a reference, the data for crystalline 4$d$ metals \cite{MTrule} and Tc-based binary hcp alloys \cite{Tcbasedalloys} are also included.
Here we define VEC = $\sum_{i=1}^{n}$$c_{i}(\rm VEC)$$_{i}$, where $c_{i}$ and (VEC)$_{i}$ are the molar fraction and valence electron number for the $i$th element, respectively.
As can be seen, the VECs of superconducting Nb-Mo-Ru-Rh-Pd HEAs fall in the range between 7.1 and 7.3, which is nearly the same as that of superconducting Ti-Mo-Re-Ru-Rh HEAs. Furthermore, in both cases, $T_{\rm c}$ tends to decrease with the increase of VEC. This trend is similar to those observed in crystalline 4d metals \cite{MTrule} and Tc-based hcp binary alloys \cite{Tcbasedalloys}, suggesting that both HEAs obey the Matthias rule.
Hence $T_{c}$ of the Nb-Mo-Ru-Rh-Pd HEAs would be maximized by reducing the VEC around 7. Nevertheless, attempts to increase the Nb content to above 20 at.\% in the Nb$_{10+2x}$Mo$_{35-x}$Ru$_{35-x}$Rh$_{10}$Pd$_{10}$ alloys   resulted in multiphase samples. In this respect, one way to lower the VEC is to replace Nb with group IVB elements Ti, Zr and Hf.

In summary, we have studied the crystal structure and physical properties of the Nb$_{10+2x}$Mo$_{35-x}$Ru$_{35-x}$Rh$_{10}$Pd$_{10}$ HEAs for $x$ in the range between 0 and 5.
The results show that all these HEAs have the hcp structure and display bulk superconductivity. With increasing Nb content $x$, $T_{\rm c}$ show a nonmonotonic behavior with a maximum of 6.19 K at $x$ = 2.5, while $B_{\rm c2}$(0) increases monotonically to 8.3 T at $x$ = 5. Both these values are the highest among hexagonal HEA superconductors. Meanwhile, the analysis of the specific heat data suggest that these HEAs have a non-BCS-like gap and their $T_{\rm c}$ is not determined primarily by the density of states at the Fermi level. In addition, $T_{\rm c}$ for these HEAs tends to decrease with increasing VEC from 7.1 to 7.3. This follows the Matthias rule for crystalline transition metal elements as well as binary alloys, and implies that $T_{\rm c}$ can be increased by lowering the valence electron number.
Our results not only identify Nb-Mo-Ru-Rh-Pd as a new element combination to form hexagonal HEAs, but also help to better understanding the superconductivity in HEAs of this structural type.
\section*{ACKNOWLEGEMENT}
This work is supported by the foundation of Westlake University and National Key Research Development Program of China (No.2017YFA0303002).

\end{document}